\documentclass[pra,twocolumn,floatfix,a4paper,superscriptaddress]{revtex4}
\usepackage{bm,color,graphicx,amsmath,txfonts}
\usepackage{here}
\usepackage{array}
\usepackage{graphicx}
\usepackage[colorlinks=true,
citecolor=blue,
linkcolor=blue,
urlcolor=blue]{hyperref}
\usepackage{braket}
\usepackage{dsfont}
\usepackage[left=	2cm,top=2cm,right=1.2cm,bottom=2cm]{geometry}
\usepackage{tikz}
\usetikzlibrary{decorations.pathmorphing, positioning}
\DeclareMathOperator{\sech}{sech}
\DeclareMathOperator{\csch}{csch}
\usepackage{tikz}
\everymath{\displaystyle}

\makeatletter
\renewcommand{\fnum@figure}{\textbf{Fig.~\thefigure:}}
\makeatother

\makeatletter
\renewcommand{\@makecaption}[2]{%
	\textbf{#1} #2\par
}
\makeatother
\makeatletter
\renewcommand{\fnum@figure}{\normalsize\textbf{Fig.~\thefigure:}}
\makeatother
\makeatletter
\renewcommand{\@makecaption}[2]{%
	\begin{flushleft}
		\textbf{#1} #2
	\end{flushleft}
}
\makeatother

\begin{document}
\makeatletter
\renewcommand{\@biblabel}[1]{%
	\makebox[2.1em][l]{\fontsize{10}{13}\selectfont[#1]}}
\makeatother

\title{Multiparameter Quantum Estimation in a Raman-Coupled Two-Qubit System}

\author{Omar Bachain}
\address{LPHE-Modeling and Simulation, Faculty of Sciences, Mohammed V University in Rabat, Rabat, Morocco}

\author{Mohamed \surname{Amazioug} }
\email{m.amazioug@uiz.ac.ma}
\address{LPTHE-Department of Physics, Faculty of Sciences, Ibnou Zohr University, Agadir 80000, Morocco}

\author{Rachid Ahl Laamara}
\address{LPHE-Modeling and Simulation, Faculty of Sciences, Mohammed V University in Rabat, Rabat, Morocco}
\address{Centre of Physics and Mathematics, CPM, Faculty of Sciences, Mohammed V University in Rabat, Rabat, Morocco}
\date{\today}
\begin{abstract}
	We investigate multiparameter quantum estimation in a Raman-coupled two-qubit system at thermal equilibrium. Analytical expressions for the quantum Fisher information matrix are derived to characterize the simultaneous estimation of the temperature and Raman coupling strength. The corresponding quantum Cram\'er--Rao bounds are obtained and compared with those of individual estimation strategies. Our results reveal optimal operating regimes determined by the interplay between thermal fluctuations and coherent interactions. In particular, quantum thermometry exhibits a well-defined optimal temperature window, whereas the estimation of the Raman coupling strength is significantly enhanced in the low-temperature and weak-coupling regime. We further show that simultaneous estimation can outperform independent estimation within appropriate parameter regions, highlighting the advantages of multiparameter quantum metrology. These results provide analytical insights into the ultimate precision limits of Raman-coupled two-qubit systems and identify promising operating regimes for quantum sensing and quantum thermometry.
\end{abstract}

\maketitle
\section{Introduction}

Quantum metrology exploits quantum mechanical effects to estimate physical parameters with a precision surpassing the limits achievable by classical strategies~\cite{Paris2009,Braunstein1994,Giovannetti2006,Giovannetti2011,Pezze2018}. At the heart of this framework lies the quantum Fisher information (QFI), which determines the ultimate precision bound through the quantum Cram\'er--Rao inequality~\cite{Helstrom1976,Holevo2001,Liu2020}. Owing to its fundamental role in parameter estimation, the QFI has become one of the cornerstones of quantum sensing, quantum imaging, atomic clocks, gravitational-wave detection, and quantum thermodynamics. Recent advances in many-body quantum systems, engineered quantum platforms, and quantum critical phenomena have further demonstrated that collective interactions, criticality, and energy-gap engineering can substantially enhance metrological sensitivity~\cite{Montenegro2025,Ye2024,ManyBodyMetrology2025}.

Beyond single-parameter estimation, multiparameter quantum metrology has attracted considerable attention because practical quantum technologies generally require the simultaneous estimation of several unknown quantities~\cite{Ragy2016,Crowley2014,Szczykulska2016}. In this scenario, the attainable precision is governed by the Quantum Fisher Information Matrix (QFIM), while quantum incompatibility and measurement-induced trade-offs introduce challenges that are absent in single-parameter estimation~\cite{Helstrom1976,Holevo2001}. More recently, multiparameter estimation has also been investigated in noisy intermediate-scale quantum (NISQ) devices, where decoherence and environmental effects strongly influence the attainable precision limits~\cite{Jiao2023}. Consequently, identifying robust quantum resources capable of improving estimation precision under realistic conditions has become one of the central objectives of modern quantum metrology.

Among the broad range of metrological applications, quantum thermometry has emerged as an important research direction in which temperature is estimated using quantum probes operating at the nanoscale~\cite{Correa2015,Mehboudi2015}. Quantum-enhanced thermometry has found applications in condensed-matter physics, quantum many-body systems, biological systems, and quantum technologies, where thermal fluctuations, quantum coherence, and interaction-induced effects jointly determine the achievable precision. Several studies have shown that optimal temperature sensitivity strongly depends on the underlying energy spectrum, the coupling between subsystems, and the operating thermal regime.

Quantum coherence has been recognized as one of the fundamental quantum resources underlying enhanced metrological performance. Since the resource-theoretic framework introduced by Baumgratz \textit{et al.}~\cite{Baumgratz2014}, quantum coherence has been extensively investigated in quantum information processing, quantum thermodynamics, quantum communication, and quantum metrology~\cite{Streltsov2017,Hu2018,Girolami2014,Scandi2025}. Recent studies have demonstrated that coherence, together with interaction-induced quantum effects, can significantly improve estimation precision even in the absence of strong entanglement, making it an important resource for realistic quantum sensing protocols.

Despite these remarkable advances, most previous investigations have focused either on single-parameter estimation or on idealized models. The simultaneous estimation of multiple physically relevant parameters in interacting quantum systems remains comparatively unexplored, particularly in molecular and spin-based platforms where thermal effects and coherent interactions coexist. Developing analytical approaches capable of characterizing the attainable precision limits in such systems is therefore of considerable interest for both fundamental quantum metrology and future quantum technologies.

Motivated by these considerations, we investigate multiparameter quantum estimation in a Raman-coupled two-qubit system at thermal equilibrium. This model provides a minimal yet nontrivial platform in which coherent interactions, thermal fluctuations, and parameter correlations coexist. We derive analytical expressions for the Quantum Fisher Information Matrix associated with the temperature and Raman coupling strength and determine the corresponding quantum Cramér--Rao bounds. A vectorization approach is employed to obtain compact analytical expressions, allowing a direct comparison between simultaneous and individual estimation strategies.

Our analysis demonstrates that the attainable precision is strongly influenced by the interplay between temperature, Raman coupling, and the intrinsic interactions of the two-qubit system. We identify the parameter regimes in which simultaneous estimation provides a clear advantage over independent estimation and analyze how thermal fluctuations affect the achievable precision limits. The analytical results obtained in this work provide useful guidelines for optimizing multiparameter quantum estimation in realistic interacting quantum systems.

The remainder of the paper is organized as follows. Section~\ref{secII} introduces the Raman-coupled two-qubit model. Section~\ref{secIII} presents the multiparameter quantum estimation framework and derives the corresponding Quantum Fisher Information Matrix. Section~\ref{secIV} discusses the estimation performance and compares simultaneous and individual estimation strategies.  Finally, Section~\ref{secVI} summarizes our main conclusions.
\section{Microscopic Model}\label{secII}

\begin{figure}[t]
	\centering
	\begin{tikzpicture}[scale=0.8]
		\draw[thick, fill=blue!30,,line width=1.2] (-3,0) circle (1.2);
		\draw[thick,line width=1.2] (-3,0.5) -- (-3,0.5);
		\draw[thick,line width=1.2] (-3,0.6) -- (-3,0.6);
		\draw[thick,line width=1.2] (-3-0.7,0.5) -- (-3+0.7,0.5);
		\draw[thick,line width=1.2] (-3-0.7,-0.5) -- (-3+0.7,-0.5);
		\draw[<->, thick,line width=1.2, red] (-3,-0.4) -- (-3,0.4);
		\draw[<->, thick,line width=1.2, red] (-3,0.4) -- (-3,-0.4);
		\node at (-2.5,0) {$\boldsymbol\omega_1$};
		\node[below=1.2cm, font=\bfseries] at (-3,0) {Left-qubit};
		\draw[thick, fill=green!30,line width=1.2] (3,0) circle (1.2);
		\draw[thick,line width=1.2] (3-0.7,0.5) -- (3+0.7,0.5);
		\draw[thick,line width=1.2] (3-0.7,-0.5) -- (3+0.7,-0.5);
		\draw[<->, thick, blue,line width=1.2] (3,-0.4) -- (3,0.4);
		\draw[<->, thick, blue,line width=1.2] (3,0.4) -- (3,-0.4);
		\node at (3.5,0) {$\boldsymbol\omega_2$};
		\node[below=1.2cm, font=\bfseries] at (3,0) {Right-qubit};
		\draw[thick, decorate, decoration={snake, amplitude=8pt, segment length=10pt}]
		(-1.8,0) -- (1.8,0);
		\node at (0,0.9) {\fontsize{9}{11}\selectfont $\bm{\mathcal{J}\left( \tau_z \otimes \tau_x \right)}$};
	\end{tikzpicture}
\caption{
	Schematic illustration of a Raman-coupled two-qubit system, in which the transition frequency of the right qubit is externally tuned, whereas the left qubit remains fixed.
}\label{fig1}
\end{figure}
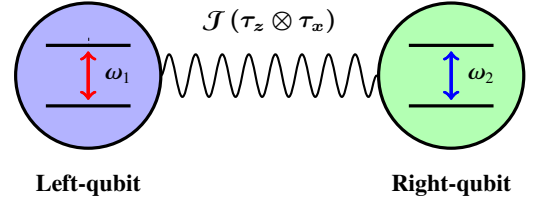

We consider a bipartite quantum system composed of two interacting qubits with transition frequencies $\omega_1$ and $\omega_2$. The interaction between the two subsystems is governed by a Raman-type anisotropic coupling \cite{Kargi2019}, as schematically illustrated in Fig.~\ref{fig1}, which induces an intrinsic asymmetry in the hybridization of the energy levels. Throughout this work, we set $\hbar = 1$.
The total Hamiltonian of the system is given by
\begin{equation}
	\mathcal{H} = \frac{\omega_1}{2} (\tau_z \otimes I) + \frac{\omega_2}{2} (I \otimes \tau_z) + \mathcal{J} (\tau_z \otimes \tau_x),
\end{equation}
where $\tau_x$ and $\tau_z$ denote the Pauli operators, $I$ is the identity operator, and $\mathcal{J}$ represents the strength of the Raman-induced coupling. This interaction differs from conventional exchange models as it couples different spin components of the two qubits, leading to a directional mixing mechanism.

In the computational basis $\{ |00\rangle, |01\rangle, |10\rangle, |11\rangle \}$, the Hamiltonian can be expressed in a block-diagonal structure, separating the dynamics into two independent subspaces.
 The diagonalization of the Hamiltonian yields four nondegenerate eigenvalues,
\begin{align*}
	\mathcal{E}_1 &= \frac{\omega_1 + \Omega}{2},\,\,\, \qquad
	\mathcal{E}_2 = \frac{\omega_1 - \Omega}{2}, \\
	\mathcal{E}_3 &= \frac{-\omega_1 + \Omega}{2}, \qquad
	\mathcal{E}_4 = \frac{-\omega_1 - \Omega}{2},
\end{align*}
where
\begin{equation}
	\Omega = \sqrt{4\mathcal{J}^2 + \omega_2^2}.
\end{equation}
The associated eigenvectors are coherent superpositions of the computational basis states and can be written as
\begin{align}\nonumber
	|\phi_1\rangle &= \cos\left( \frac{\varphi}{2}\right)  |00\rangle + \sin\left( \frac{\varphi}{2} \right) |01\rangle, \\\nonumber
	|\phi_2\rangle &= \sin\left( \frac{\varphi}{2}\right)  |00\rangle - \cos\left( \frac{\varphi}{2} \right) |01\rangle, \\\nonumber
	|\phi_3\rangle &= \sin\left( \frac{\varphi}{2}\right)  |10\rangle - \cos\left( \frac{\varphi}{2}\right)  |11\rangle, \\
	|\phi_4\rangle &= \cos\left( \frac{\varphi}{2}\right)  |10\rangle + \sin\left( \frac{\varphi}{2}\right)  |11\rangle,
\end{align}
where the mixing angle $\varphi$ satisfies
\begin{equation}
	\tan(\varphi) = \frac{2\mathcal{J}}{\omega_2}.
\end{equation}

At thermal equilibrium with a reservoir at temperature $T$, the system is described by the Gibbs state
\begin{equation}
	\varrho(T) = \frac{e^{-\beta \,\mathcal{H}}}{Z} = \sum_{i=1}^{4} p_i |\phi_i\rangle \langle \phi_i|,
\end{equation}
where $\beta = 1/(k_B T)$ and the occupation probabilities are given by
\begin{equation}
	p_i = \frac{e^{-\beta\, \mathcal{E}_i}}{Z}, \qquad Z = \sum_{i=1}^{4} e^{-\beta \,\mathcal{E}_i}.
\end{equation}
Due to the block-diagonal structure of the Hamiltonian, the density matrix preserves the same form in the computational basis and can be written as
\begin{equation}
	\rho(T) = \frac{1}{Z}
	\begin{pmatrix}
		\varrho^{+} & 0 \\
		0 & \varrho^{-}
	\end{pmatrix},
\end{equation}
where $\varrho^{+}$ and $\varrho^{-}$ are $2 \times 2$ matrices acting on the subspaces $\{ |00\rangle, |01\rangle \}$ and $\{ |10\rangle, |11\rangle \}$, respectively.
The elements of $\varrho^{+}$ are explicitly given by
\begin{align}\nonumber
	\varrho^{+}_{11} &= e^{-\beta\, \omega_1/2} \left[\cosh\left(\frac{\beta\, \Omega}{2}\right) - \frac{\omega_2}{\Omega}\sinh\left(\frac{\beta \Omega}{2}\right)\right], \\\nonumber
	\varrho^{+}_{22} &= e^{-\beta\, \omega_1/2} \left[\cosh\left(\frac{\beta \,\Omega}{2}\right) + \frac{\omega_2}{\Omega}\sinh\left(\frac{\beta \Omega}{2}\right)\right], \\
	\varrho^{+}_{12} &= \rho^{+}_{21} = -\frac{2\mathcal{J}}{\Omega} e^{-\beta \,\omega_1/2} \sinh\left(\frac{\beta\, \Omega}{2}\right),
\end{align}
while the elements of $\rho^{-}$ read
\begin{align}\nonumber
	\varrho^{-}_{11} &= e^{\beta \,\omega_1/2} \left[\cosh\left(\frac{\beta \,\Omega}{2}\right) - \frac{\omega_2}{\Omega}\sinh\left(\frac{\beta \Omega}{2}\right)\right], \\\nonumber
	\varrho^{-}_{22} &= e^{\beta \,\omega_1/2} \left[\cosh\left(\frac{\beta\, \Omega}{2}\right) + \frac{\omega_2}{\Omega}\sinh\left(\frac{\beta \Omega}{2}\right)\right], \\
	\varrho^{-}_{12} &= \rho^{-}_{21} = \frac{2\mathcal{J}}{\Omega} e^{\beta \,\omega_1/2} \sinh\left(\frac{\beta\, \Omega}{2}\right).
\end{align}
Accordingly, the full $4 \times 4$ density matrix takes the explicit form
\begin{equation}
	\varrho=
	\begin{pmatrix}
		\varrho^+_{11}& \varrho^+_{12} & 0 & 0 \\
		\varrho^+_{21} &\varrho^+_{22} & 0 & 0 \\
		0 & 0  & \varrho^-_{11} & \varrho^-_{12} \\
		0 & 0 & \varrho^-_{21} & \varrho^-_{22}
	\end{pmatrix}.\label{densit M}
\end{equation}

The presence of nonzero off-diagonal terms reflects the emergence of quantum coherence induced by the Raman coupling $\mathcal{J}$. In the limit $\mathcal{J} \rightarrow 0$, the mixing vanishes and the density matrix reduces to a diagonal form, corresponding to a classical statistical mixture of energy eigenstates.
\section{QUANTUM FISHER INFORMATION MATRIX}    \label{secIII}

In this section, we establish the framework of multiparameter quantum estimation and derive a compact analytical expression of the quantum Fisher information matrix (QFIM) based on a density-operator vectorization approach. This method avoids the explicit diagonalization of the density matrix and provides an efficient computational scheme for arbitrary finite-dimensional quantum systems.
Let $\mathcal{H}$ be an $n$-dimensional Hilbert space and $\mathcal{B}(\mathcal{H})$ the set of linear operators acting on it. For any operator $A \in \mathcal{B}(\mathcal{H})$, we define its vectorized form using Dirac notation as \cite{gilchrist2009vectorization}
\begin{equation}
	|A\rangle\rangle = \sum_{k,l} a_{kl} \, |k\rangle \otimes |l\rangle .
\end{equation}
In particular, for a matrix
\begin{equation}
	A =
	\begin{pmatrix}
		a_{11} & a_{12} & \cdots & a_{1n} \\
		a_{21} & a_{22} & \cdots & a_{2n} \\
		\vdots & \vdots & \ddots & \vdots \\
		a_{n1} & a_{n2} & \cdots & a_{nn}
	\end{pmatrix},
\end{equation}
the vectorization corresponds to stacking the columns of $A$ into a single vector,
\begin{equation}
	|A\rangle\rangle = (a_{11}, a_{21}, \ldots, a_{n1}, a_{12}, \ldots, a_{nn})^T.
\end{equation}

Equivalently, the vectorization map can be written as
\begin{equation}
	|A\rangle\rangle = (I \otimes A) \sum_{i=1}^{n} |i\rangle \otimes |i\rangle .
\end{equation}
The following identities hold:
\begin{align}
	(A \otimes B)|C\rangle\rangle &= |ACB^{T}\rangle\rangle, \\
	\langle\langle A|B\rangle\rangle &= \mathrm{Tr}(A^\dagger B), \\
	|A\rangle\rangle &= \sum_{k,l} a_{kl} |k\rangle |l\rangle .
\end{align}

We consider a density operator $\varrho(\boldsymbol{\lambda})$ depending smoothly on parameters
\[
\boldsymbol{\lambda}=(\lambda_1,\ldots,\lambda_m).
\]
The covariance matrix of any unbiased estimator satisfies the quantum Cram\'er–Rao bound \cite{Paris2009}
\begin{equation}
	\mathrm{Cov}(\hat{\boldsymbol{\lambda}})\ge \mathcal{I}^{-1}.
\end{equation}

The QFIM elements are
\begin{equation}
	\mathcal{I}_{\mu\nu}
	=
	\frac{1}{2}
	\mathrm{Tr}\!\left[
	\varrho({\mathcal{L}}_\mu\mathcal{L}_\nu+{\mathcal{L}}_\nu\mathcal{\mathcal{L}}_\mu)
	\right].
\end{equation}

The SLD operators satisfy
\begin{equation}
	\partial_{\lambda_\mu}\varrho
	=
	\frac{1}{2}({\mathcal{L}}_\mu\varrho+\varrho{\mathcal{L}}_\mu).
\end{equation}

For a single parameter \cite{helstrom1969quantum,Holevo2001}
\begin{equation}
	\left( \Delta\lambda\right)^2 \ge \frac{1}{\mathcal{I}(\lambda)}, \quad 
	\mathcal{I}(\lambda)=\mathrm{Tr}[\varrho {\mathcal{L}}_\lambda^2].
\end{equation}

Using spectral decomposition \cite{banchi2014quantum,Sommers2003}
\begin{equation}
	\mathcal{I}_{\mu\nu}
	=
	2\sum_{r_k+r_l>0}
	\frac{\langle k|\partial_{\lambda_\mu}\varrho|l\rangle
		\langle l|\partial_{\lambda_\nu}\varrho|k\rangle}{r_k+r_l}.
\end{equation}

To avoid diagonalization, define \cite{vsafranek2018simple,Matsumoto2002,crowley2014tradeoff}
\begin{equation}
	\Lambda=\varrho^T\otimes I + I\otimes\varrho.\label{nabla}
\end{equation}

Then
\begin{equation}
	\mathcal{I}_{\mu\nu}
	=
	2\,|\partial_{\lambda_\mu}\varrho\rangle\rangle^T
	\Lambda^{-1}
|\partial_{\lambda_\nu}\varrho\rangle\rangle.\label{F}
\end{equation}
The SLDs are
\begin{equation}
	|{L}_\mu\rangle\rangle=2\Lambda^{-1}|\partial_{\lambda_\mu}\varrho\rangle\rangle.\label{SLD}
\end{equation}

The vectorized derivatives are
\begin{align}\nonumber
	|\partial_T \varrho\rangle\rangle=
	\Big(\partial_T \varrho^+_{11},\,\partial_T\varrho^+_{21},\,0\,,0\,,\,\partial_T\varrho^+_{12},\partial_T\varrho^+_{22},\,0,\,&0,\,0,\,0,\,\\\nonumber
	\partial_T\varrho^-_{11},\,\partial_T\varrho^-_{21},\,0,\,0,\,\partial_T\varrho^+_{12},\,\partial_T\varrho^+_{22}	\Big),\\\nonumber
	|\partial_\mathcal{J} \varrho\rangle\rangle=	
	\Big(\partial_\mathcal{J} \varrho^+_{11},\,\partial_\mathcal{J}\varrho^+_{21},\,0\,,0\,,\,\partial_\mathcal{J}\varrho^+_{12},\,\partial_\mathcal{J}\varrho^+_{22},\,0,\,&0,\,0,\,0,\,\\
	\partial_\mathcal{J}\varrho^-_{11},\,\partial_\mathcal{J}\varrho^-_{21},\,0,\,0,\,\partial_\mathcal{J}\varrho^+_{12},\,\partial_\mathcal{J}\varrho^+_{22}	\Big).
\end{align}

Thus the QFIM becomes
\begin{align*}
	\mathcal{I}&=
	\begin{pmatrix}
		\mathcal{I}_{TT}&\mathcal{I}_{T\mathcal{J}}\\
		\mathcal{I}_{\mathcal{J}T}&\mathcal{I}_{\mathcal{J}\mathcal{J}}
	\end{pmatrix}\\
	&=\begin{pmatrix}
	2\,|\partial_{T} \varrho\rangle\rangle^T 
	\Lambda^{-1} 
	|\partial_{T} \varrho \rangle\rangle
	&
	2\,|\partial_{T} \varrho\rangle\rangle^T 
	\Lambda^{-1} 
	|\partial_\mathcal{J} \varrho\rangle\rangle
	\\[6pt]
	2\,|\partial_\mathcal{J} \varrho\rangle\rangle^T 
	\Lambda^{-1} 
	|\partial_{T} \varrho\rangle\rangle
	&
	2\,|\partial_\mathcal{J} \varrho\rangle\rangle^T 
	\Lambda^{-1} 
	|\partial_\mathcal{J} \varrho\rangle\rangle
	\end{pmatrix},
\end{align*}
where 
\begin{align}\nonumber
	\mathcal{I}_{TT}
	&=
	\frac{
		\omega_1^2 \, \mathrm{sech}^2\!\left(\frac{\omega_1}{2T}\right)
		+ \Omega^2 \, \mathrm{sech}^2\!\left(\frac{\Omega}{2T}\right)
	}{
		4T^4
	},\\\nonumber
	\mathcal{I}_{\mathcal{J}\mathcal{J}}
	&=
	\frac{
		4 T^2 \omega_2^2 
		+ 4 \left(4 \mathcal{J}^4 + (\mathcal{J}^2 - T^2)\, \omega_2^2 \right)
		\, \mathrm{sech}^2\!\left(\frac{\Omega}{2T}\right)
	}{
		T^2 \Omega^4
	},\\
	\mathcal{I}_{\mathcal{J} T}
	&=\mathcal{I}_{T\mathcal{J} }
	=
	-\frac{
		\mathcal{J} \, \mathrm{sech}^2\!\left(\frac{\Omega}{2T}\right)
	}{
		T^3
	}.
\end{align}

The quantum Cram\'er–Rao bound gives \cite{Prussing1986}
\begin{equation}
	\left(\Delta T \right)^2 \ge \frac{\mathcal{I}_{\mathcal{J}\mathcal{J}}}{\det\mathcal{I}},\quad
	\left( \Delta \mathcal{J}\right) ^2\ge \frac{\mathcal{I}_{TT}}{\det\mathcal{I}}.
\end{equation}
The minimal values of $\left(\Delta T \right)^2 $ and $\left(\Delta \mathcal{J} \right)^2 $ are given by
\begin{align}
	\left(\Delta T \right)^2_{\min} &=
		\frac{
			8 T^6 \omega_2^2 
			+ 8 T^4 \left(4 \mathcal{J}^4 + (\mathcal{J}^2-T^2)\omega_2^2\right)
			\sech^2\!\left(\frac{\Omega}{2T}\right)
		}{D_1},
\end{align}
\begin{align}\nonumber
	D_1&=
		\omega_1^2 
		\Bigg[
		8 \mathcal{J}^4 
		+ (2\mathcal{J}^2 - T^2)\omega_2^2 
		+ T^2 \omega_2^2 \cosh\!\left(\frac{\Omega}{T}\right)
		\Bigg]
		\sech^2\!\left(\frac{\omega_1}{2T}\right)\\
		&\times\sech^2\!\left(\frac{\Omega}{2T}\right)
		+32 T^2 \omega_2^2 \Omega^2
		\csch^4\!\left(\frac{\Omega}{T}\right)
		\sinh^6\!\left(\frac{\Omega}{2T}\right),
\end{align}
and
\begin{align}
	\left(\Delta \mathcal{J} \right)^2_{\min} =	\frac{T^2 \Omega^4 
	\left[
	\omega_1^2 \sech^2\!\left(\frac{\omega_1}{2T}\right)
	+ \Omega^2 \sech^2\!\left(\frac{\Omega}{2T}\right)
	\right]}{D_2},
\end{align}
\begin{align}\nonumber
	D_2&= 
	2 \omega_1^2 
	\left(
	8 \mathcal{J}^4 
	+ (2\mathcal{J}^2 - T^2)\omega_2^2
	+ T^2 \omega_2^2 \cosh\!\left(\frac{\Omega}{T}\right)
	\right)
	\\\nonumber
	&\quad \times 
	\sech^2\!\left(\frac{\omega_1}{2T}\right)\sech^2\!\left(\frac{\Omega}{2T}\right) + 
	64 T^2 \omega_2^2 \Omega^2 
	\csch^4\!\left(\frac{\Omega}{T}\right)\\
	&\quad \times
	\sinh^6\!\left(\frac{\Omega}{2T}\right)
\end{align}
The SLDs are
\begin{equation}
	\mathcal{L}_T =
	\begin{pmatrix}
		A_1 & \frac{\mathcal{J}}{T^2} & 0 & 0 \\
		\frac{\mathcal{J}}{T^2} & A_2 & 0 & 0 \\
		0 & 0 & A_3 &\frac{\mathcal{J}}{T^2} \\
		0 & 0 & \frac{\mathcal{J}}{T^2} & A_4
	\end{pmatrix},
\end{equation}
where 
\begin{align}\nonumber
	A_1 &= \frac{
		\left(2 - \frac{2}{1+e^{\omega_1/T}}\right)\omega_1
		+ \omega_2
		+ \Omega \tanh\!\left(\frac{\Omega}{2T}\right)
	}{2T^2}, \\\nonumber
	A_2 &= -\frac{
		2\left(-1 + \frac{1}{1+e^{\omega_1/T}}\right)\omega_1
		+ \omega_2
		- \Omega \tanh\!\left(\frac{\Omega}{2T}\right)
	}{2T^2}, \\\nonumber
	A_3 &= \frac{
		-\frac{2\omega_1}{1+e^{\omega_1/T}}
		+ \omega_2
		+ \Omega \tanh\!\left(\frac{\Omega}{2T}\right)
	}{2T^2}, \\
	A_4 &= -\frac{
		\frac{2\omega_1}{1+e^{\omega_1/T}}
		+ \omega_2
		- \Omega \tanh\!\left(\frac{\Omega}{2T}\right)
	}{2T^2},
\end{align} 
and
\begin{equation}
	L_{\mathcal{J}} =
	\begin{pmatrix}
		C_1 & D & 0 & 0 \\
		D & C_2 & 0 & 0 \\
		0 & 0 & C_1 & D \\
		0 & 0 & D & C_3
	\end{pmatrix},
\end{equation}
with
\begin{align}\nonumber
	C_1 &= -\frac{
		2\mathcal{J}\Big[
		\omega_2 \Omega
		+ \left(4\mathcal{J}^2 + \omega_2(\omega_2 - 2T)\right)
		\tanh\!\left(\frac{\Omega}{2T}\right)
		\Big]
	}{
		T\, \Omega^3
	}, 	\\\nonumber
	C_2 &= \frac{
		2\mathcal{J}\,\omega_2\,\Omega
		-2\mathcal{J}\left(4\mathcal{J}^2 + \omega_2(\omega_2 + 2T)\right)
		\tanh\!\left(\frac{\Omega}{2T}\right)
	}{
		T\, \Omega^3
	}, \\\nonumber
	C_3 &= \frac{
		e^{-\omega_1/T}(1+e^{\omega_1/T})\,\mathcal{J}
		\left(1+\tanh\!\frac{\omega_1}{2T}\right)
	}{
		T\, \Omega^4
	}
	\times
	\Bigg[
	4\mathcal{J}^2 \omega_2 + \omega_2^3\\\nonumber
	&- \Omega \left(4\mathcal{J}^2 + \omega_2(\omega_2+2T)\right)
	\tanh\!\left(\frac{\Omega}{2T}\right)
	\Bigg],\\
	D&=-\frac{4\mathcal{J}^2}{T(4\mathcal{J}^2+\omega_2^2)}
	-\frac{2\omega_2^2}{\Omega^3}
	\tanh\!\left(\frac{\Omega}{2T}\right).
\end{align}

To analyze the attainability of the quantum Cram\'er--Rao bound, we examine the commutation relation between these operators. In general, \(\mathcal{L}_T\) and \(\mathcal{L}_{\mathcal{J}}\) do not commute, which prevents the construction of a common eigenbasis. However, commutativity is only a sufficient condition, not a necessary one.
A weaker but sufficient condition for the saturation of the multiparameter quantum Cram\'er--Rao bound is given by
\begin{equation}
	\mathrm{Tr}\big(\rho\, [\mathcal{L}_T, \mathcal{L}_{\mathcal{J}}]\big) = 0,
\end{equation}
which ensures the asymptotic attainability of the bound. This condition expresses a form of statistical compatibility between the parameters, even in the absence of commuting SLDs.
Therefore, by evaluating this trace condition using the explicit forms of \(\mathcal{L}_T\) and \(\mathcal{L}_{\mathcal{J}}\), one can determine whether the simultaneous estimation of \(T\) and \(\mathcal{J}\) can achieve the ultimate precision limit. This provides deeper insight into the interplay between thermal effects and interaction-induced correlations in quantum parameter estimation\cite{ragy2016compatibility,vrehavcek2018optimal}.
\begin{figure}
	\includegraphics[width=0.9\linewidth]{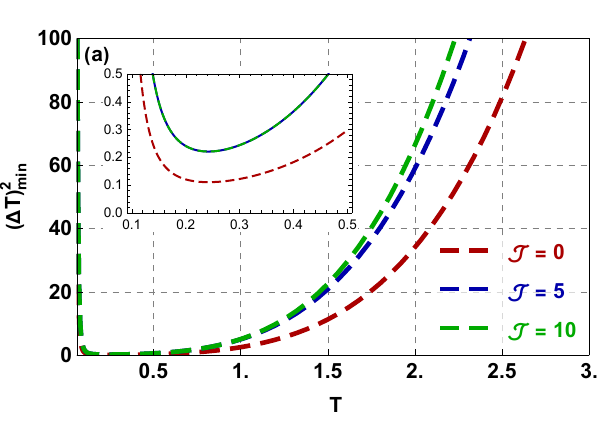}
	\includegraphics[width=0.9\linewidth]{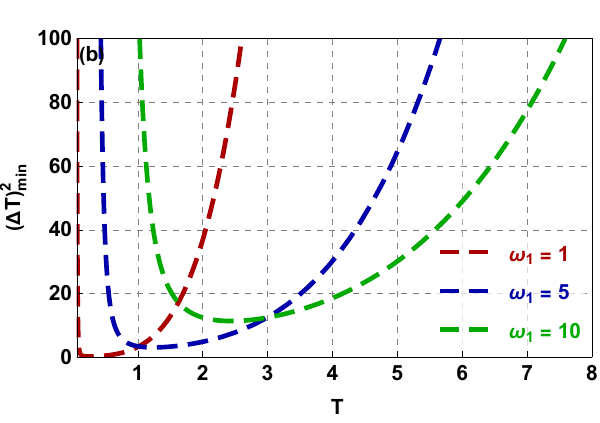}
	\caption{Minimum variance $(\Delta T)_{\min}^2$ as a function of temperature $T$. (a) For different values of the coupling strength $\mathcal{J}$ with $\omega_1=\omega_2=1$. (b) For different values of the transition frequency $\omega_1=\omega_2$ with $\mathcal{J}=1$.}
	\label{fig2}
\end{figure}
\begin{figure}
	\includegraphics[width=0.9\linewidth]{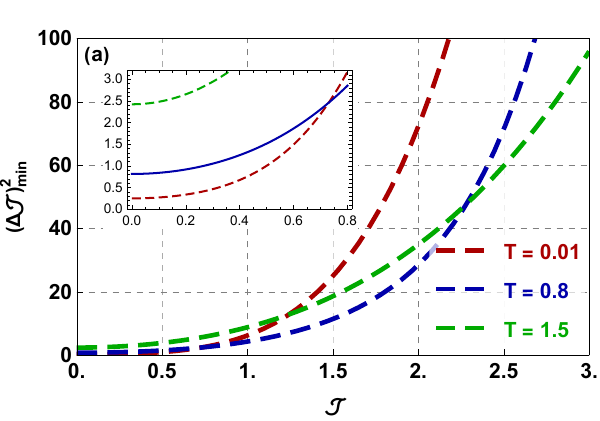}
	\includegraphics[width=0.9\linewidth]{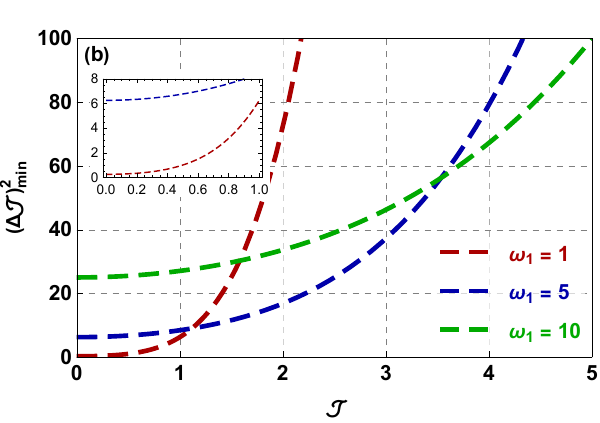}
	\caption{Minimum variance $(\Delta \mathcal{J})_{\min}^2$ as a function of the coupling strength $\mathcal{J}$. (a) For different values of temperature $T$ with $\omega_1=\omega_2=1$. (b) For different values of the transition frequency $\omega_1=\omega_2$ with $T=0.01$.}
	\label{fig3}
\end{figure}

The behavior of the minimum temperature variance $(\Delta T)^2_{\min}$ as a function of temperature is shown in Fig.~\ref{fig2}.
In both panels, the variance exhibits a non-monotonic behavior as a function of temperature. In the low-temperature limit ($T \to 0$), $(\Delta T)^2_{\min}$ becomes very large. This divergence originates from the fact that the system approaches its ground state, making it almost insensitive to temperature variations and thus leading to poor estimation precision.
As the temperature increases, thermal excitations progressively populate higher energy levels, enhancing the distinguishability between neighboring states. As a result, the variance decreases and reaches a minimum within an intermediate temperature regime, defining an optimal working region for quantum thermometry. Beyond this regime, at higher temperatures, thermal fluctuations dominate and degrade the estimation precision, causing $(\Delta T)^2_{\min}$ to increase again.
For Fig.~\ref{fig2}(a), corresponding to different values of the coupling strength $\mathcal{J}$ with $\omega_1=\omega_2=1$, the optimal region is essentially independent of $\mathcal{J}$, with $(\Delta T)^2_{\min}<1$ for $T \in [0.12,\,0.61]$. This indicates that the coupling has a limited influence on the temperature sensitivity in this regime.
In contrast, Fig.~\ref{fig2}(b) shows that the optimal temperature window strongly depends on the transition frequencies $\omega_1=\omega_2$ at fixed $\mathcal{J}=1$. For instance, for $\omega_1=1$, the condition $(\Delta T)^2_{\min}<1$ holds approximately for $T \in [0.13,\,0.69]$. Increasing $\omega_1$ shifts this optimal region, highlighting that the estimation precision is primarily controlled by the characteristic energy scale of the system.
Overall, these results demonstrate that while the coupling $\mathcal{J}$ has a minor effect on the optimal temperature window, the transition frequencies $\omega_i$ play a crucial role through the energy gap $\Omega$, which governs the thermal sensitivity of the system.

The behavior of the minimum variance $(\Delta \mathcal{J})^2_{\min}$ as a function of the coupling strength $\mathcal{J}$ is shown in Fig.~\ref{fig3}.
In Fig.~\ref{fig3}(a), for fixed $\omega_1=\omega_2=1$, the variance strongly depends on temperature. At very low temperature ($T=0.01$), $(\Delta \mathcal{J})^2_{\min}<1$ for $\mathcal{J} \in [0,\,0.51]$, indicating a high precision regime. As the temperature increases, this optimal region shrinks: for $T=0.8$, the condition $(\Delta \mathcal{J})^2_{\min}<1$ holds approximately for $\mathcal{J} \in [0,\,0.29]$, while for $T=1.5$, the variance remains below $3$ only up to $\mathcal{J} \approx 0.32$. This clearly shows that the estimation precision degrades with increasing temperature. Physically, this behavior arises from thermal fluctuations that progressively wash out the sensitivity of the state to variations of the coupling parameter.
In addition, for all temperatures, $(\Delta \mathcal{J})^2_{\min}$ increases rapidly with $\mathcal{J}$, indicating that large coupling strengths reduce the distinguishability between neighboring states and thus limit the estimation precision.
In Fig.~\ref{fig3}(b), at fixed temperature $T=0.01$, the variance exhibits a strong dependence on the transition frequency $\omega_1=\omega_2$. For $\omega_1=1$, a high-precision regime $(\Delta \mathcal{J})^2_{\min}<1$ is observed for $\mathcal{J} \in [0,\,0.49]$. However, for larger values of $\omega_1$, the variance becomes significantly larger, exceeding values of order $6$ and $20$ even at small $\mathcal{J}$, and continues to increase with $\mathcal{J}$. This behavior reflects the fact that increasing the energy scale suppresses the sensitivity of the system to variations of the coupling strength.
Overall, these results demonstrate that the estimation of $\mathcal{J}$ is optimal at low temperature and small coupling, while both thermal effects and large energy scales significantly degrade the precision.

\section{Conclusion}\label{secVI}

In this work, we have investigated multiparameter quantum estimation in a Raman-coupled two-qubit system at thermal equilibrium. By deriving analytical expressions for the quantum Fisher information matrix, we analyzed the simultaneous estimation of the temperature and the Raman coupling strength, and compared its performance with that of individual estimation strategies.

Our results demonstrate that the attainable precision strongly depends on the interplay between thermal fluctuations and coherent interactions. In particular, quantum thermometry exhibits a well-defined optimal temperature window, while the estimation of the Raman coupling strength is significantly enhanced in the low-temperature and weak-coupling regime. These findings highlight the fundamental role of the energy spectrum and interaction strength in determining the ultimate precision limits.

Furthermore, we identified parameter regimes in which simultaneous estimation provides a clear advantage over independent estimation, illustrating the potential of multiparameter protocols for improving the efficiency of quantum sensing. The analytical expressions obtained for the quantum Fisher information matrix also provide a transparent description of how the system parameters influence the estimation precision.

Overall, our results establish the Raman-coupled two-qubit system as a promising platform for multiparameter quantum metrology. Beyond providing analytical precision bounds, this work offers useful guidelines for optimizing quantum estimation protocols in interacting quantum systems. We expect that the proposed framework can be extended to more complex spin systems, open quantum systems, and experimentally accessible quantum platforms, thereby contributing to the development of high-precision quantum sensing and quantum thermometry.

\twocolumngrid

\end{document}